\documentclass{sig-alternate-05-2015}
\usepackage{graphicx}       
\usepackage{amsmath,amssymb} 
\usepackage{color}
\usepackage{times}
\usepackage{epsfig}
\usepackage{amsmath}
\usepackage{amssymb}
\usepackage{enumitem}
\usepackage{array}
\usepackage{float}
\usepackage{hyperref}

\begin{document}

\setcopyright{acmcopyright}

\conferenceinfo{EMPIRE 2016,} {September 16, 2016, Boston, MA, USA.}
\CopyrightYear{2016}
\newtoks\copyrightetc
\global\copyrightetc{Copyright remains with the authors and/or original copyright holders}

\doi{10.13140/RG.2.1.4167.0649}

\isbn{0000-0003-4170-914X}



%

\title{Towards Personality-Aware Recommendation}
\subtitle{Introducing ADS Dataset}
%
%
%
%
%

\numberofauthors{1} 
%
\author{
%
%
\alignauthor
Giorgio Roffo\\
       \affaddr{Department of Computer Science}\\
      \affaddr{University of Verona (IT)}\\
       \email{giorgio.roffo@univr.it}
}

\maketitle
\begin{abstract}
In the last decade new ways of shopping online have increased the possibility of buying products and services more easily and faster than ever. In this new context, personality is a key determinant in the decision making of the consumer when shopping. The two main reasons are: firstly, a person's buying choices are influenced by psychological factors like impulsiveness, and secondly, some consumers may be more susceptible to making impulse purchases than others. To the best of our knowledge, the impact of personality factors on advertisements has been largely neglected at the level of recommender systems. This work proposes a highly innovative research which uses a personality perspective to determine the unique associations among the consumer's buying tendency and advert recommendations. As a matter of fact, the lack of a publicly available benchmark for computational advertising do not allow both the exploration of this intriguing research direction and the evaluation of state-of-the-art algorithms. We present the ADS Dataset, a publicly available benchmark for computational advertising enriched with Big-Five users' personality factors and 1,200 personal users' pictures. The proposed benchmark allows two main tasks: rating prediction over 300 real advertisements (i.e., Rich Media Ads, Image Ads, Text Ads) and click-through rate prediction. Moreover, this work carries out experiments, reviews various evaluation criteria used in the literature, and provides a library for each one of them within one integrated toolbox.
\end{abstract}

%
%


%
%

%
%


\section{Introduction}\label{sec:intro}\vspace{-0.08cm}

Nowadays, online shopping plays an increasingly significant role in our daily lives~\cite{Forrester}. Most consumers shop online with the majority of these shoppers preferring to shop online for reasons like saving time and avoiding crowds. Marketing campaigns can create awareness that drive consumers all the way through the process to actually making a purchase online~\cite{Kim:2011}. Accordingly, a challenging problem is to provide the user with a list of recommended advertisements they might prefer, or predict how much they might prefer the content of each advert.

Attitudes, perceptions and motivations are not apparent from clicks on advertisements or online purchases, but they are an important part of the success or failure of online marketing strategies. A person's buying choices are further influenced by psychological factors like impulsiveness (e.g., leads to impulse buying behaviors), neuroticism, extraversion which affect their motivations and attitudes~\cite{Turkyilmaz201598}. However, the impact of personality factors on advertisements has been studied at the level of social sciences and  microeconomics~\cite{Bosnjak2007,Turkyilmaz201598,soBob}, but largely neglected, to the best of our knowledge, at the level of advert recommender systems. Taking the lack of publicly available benchmark which include questionnaires consisting of the items of validated scales (e.g., Five Factor Model~\cite{rammstedt:2007}) on the involved individuals, this study is meant to contribute to this unexplored area. 

This paper presents the ADS Dataset, a highly innovative collection of 300 real advertisements rated by 120 individuals and enriched with their personality traits. The user study is conducted by recruiting a set of test subjects, and asking them to perform several tasks. The experimental protocol adopted for the data collection has been designed to capture users' preferences in a controlled usage scenario so as to avoid any biases. Moreover, this work carries out experiments with various evaluation criteria used in the literature and provides a library within one integrated toolbox.

Summarizing, the contribution of this work is three-fold:

\textbf{Dataset}: we collect a representative benchmark for computational advertising enriched with affective-like features such as personality factors and visual cues from users' favorite pictures. The benchmark allows us to (i) explore the relationship between consumer characteristics, attitude toward online shopping and advert recommendation, (ii) identify the underlying dimensions of consumer shopping motivations and attitudes toward online in-store conversions, and (iii) have a reference benchmark for comparison of state-of-the-art advertisement recommender systems (ARSs).

\textbf{Code library}: we integrated most widely used evaluation metrics in the proposed code library with uniform input and output formats to facilitate large scale performance evaluation. 

\textbf{Evaluation metrics}: we collect two broad classes of prediction accuracy measures, depending on the task the recommender system is performing: ``ad rating prediction'' or ``ad click prediction''.

The code library, annotated dataset and all the results are available on \href{http://giorgioroffo.it/}{ADS Dataset} project page.

The rest of the paper is organized as follows: in Section~\ref{sec:dataset} we present and describe the proposed ADS Dataset. In Section~\ref{sec:Algorithms} we show the baseline techniques used for the recommendation of advertisements. Then, in Section~\ref{sec:eval}, we survey a large set of evaluation metrics in the context of the property that ARSs evaluate. In Section~\ref{sec:results} results are reported for each one of the two scenarios taken into account in this work, and finally, in Section~\ref{sec:conclusion} conclusions are given, and future perspectives are envisaged.

\begin{table*}[!ht]\vspace{-0.19cm}
\centering
  \resizebox{1\textwidth}{!}{%
\begin{tabular}{|m{3cm}|m{3.9cm}|m{15cm}|m{1.4cm}|}
\hline
\textbf{Group}  &\textbf{Type} & \textbf{Description} & \textbf{References}\\\hline
\textbf{Users' Preferences} 
& Websites, Movies, Music, TV Programmes, Books, Hobbies &  Categories of: websites users most often visit (WB), watched films (MV), listened music (MS), watched T.V. Programmes (TV), books users like to read (BK), favourite past times, kinds of sport, travel destinations.& \cite{He:2014,NzVald20121186}\\

\hline
\textbf{Demographic} & Basic information  & Age, nationality, gender, home town, CAP/zip-code, type of job, weekly working hours, monetary well-being of the participant &\cite{NzVald20121186}\\

\hline
 \textbf{Social Signals} &  Personality Traits&  BFI-10: Openness to experience, Conscientiousness, Extraversion, Agreeableness, and Neuroticism (OCEAN) &~\cite{Chen201657,rammstedt:2007}\\

\hline
 \textbf{Users' Ratings} &  Clicks & 300 ads annotated with Click / No Click by 120 subjects &~\cite{Avila16,wang2010click,He:2014}\\
 &  Feedback  & From 1-star (Negative) to 5-stars (Positive) users' feedback on 300 ads &~\cite{Avila16,wang2010click,He:2014}\\ 

\hline	
\end{tabular}}
  \caption{The table reports the type of raw data provided by the ADS Dataset. Data of the first and last group can be considered as historical information about the users in an offline user study. }
  \label{tab:features}
\end{table*}

\section{ADS Dataset}\label{sec:dataset}\vspace{-0.08cm}

The corpus includes $300$ advertisements voted by unacquainted individuals (120 subjects in total. Note, the data collection process is still running). Advert content is categorized in terms of the $20$ main  Amazon product categories. Adverts equally cover three display formats: Rich Media Ads, Image Ads, Text Ads (i.e., $100$ ads for each format). Participants rated (from 1-star to 5-stars) each recommended advertisement according to if they would or would not click on it.
Inspired from recent findings which investigate the effects of personality traits on online impulse buying~\cite{Bosnjak2007,Turkyilmaz201598,soBob}, and many other approaches based upon behavioral economics, lifestyle analysis, and merchandising effects~\cite{Bosnjak2007,Mowen2000}, this study takes a trait theory approach to studying the effect of personality on user's motivations and attitudes toward online in-store conversions. 
Therefore, the corpus includes the Big Five Inventory-10 (BFI-10) to measure personality traits~\cite{rammstedt:2007}, which allows the exploration of this promising research direction. 

Recent soft-biometric approaches have shown the ability to infer the personality traits of users from visual cues extracted from their favorite pictures (e.g., in Computational Aesthetics), or from users' writing behaviors in social media settings~\cite{Roffo:icmi2014,Roffo:HBU2014,Roffo_2013_ICCV_Workshops}. While not necessarily corresponding to the actual traits of an individual, attributed traits are still important because they are predictive of important aspects of social life, including attitude of others and social identity. As a result, the proposed benchmark also includes 1,200 spontaneously uploaded images that hold a lot of meaning for the participants and their related annotations: \textit{positive/negative} (see Table~\ref{tab:features} for further details).

\subsection{The Subjects}\label{sec:subjects}\vspace{-0.08cm}

This corpus involves 120 English native speakers between 18 and 68. The median of the participants age is 28. Most of the participants have a university education, 31\% of them are undergraduate students. In terms of gender, 77 are females and 43 males. The percentage distribution of household income within the sample is: 23\% less or equal to 11K USD per year, 48\% from 11K to 50K USD, 21\% from 50K to 85K USD, and 8\% more than 85K USD. The median income is between 11K and 50K USD. 

\section{Evaluated Algorithms}\label{sec:Algorithms}\vspace{-0.08cm}

Since a prediction engine lies at the basis of the most recommender systems, we selected some of the most widely used techniques for recommendations and predictions~\cite{He:2014}, such as Logistic Regression (LR)~\cite{LIBLINEAR}, Support Vector Regression with radial basis function (SVR-rbf)~\cite{Chang:2011}, and L2-regularized L2-loss Support Vector Regression (L2-SVR)~\cite{LIBLINEAR}. 
These engines may predict user opinions to adverts (e.g., a user's positive or negative feedback to an ad) or the probability that a user clicks or performs a conversion (e.g., an in-store purchase) when they see an ad. In Section~\ref{sec:results}, we evaluate these methods while feeding them with and without features coming from the psychometric traits. \\\\

\section{Evaluation Methodology}\label{sec:eval}\vspace{-0.08cm}

Research in the ARS field requires quality measures and evaluation metrics to know the quality of the techniques, methods, and algorithms for predictions and recommendations. In this section we review the process of evaluating an ARS on two main tasks: (i) measuring the accuracy of rating predictions, and (ii) measuring the accuracy of click predictions.

\subsection{Ad Rating Prediction}\label{sec:}\vspace{-0.08cm}

In this first scenario, we want to predict the feedback a user would give to an advert (e.g. 1-star through 5-stars). In such a case, we want to measure the accuracy of the system`s
predicted ratings. \textbf{Root Mean Squared Error (RMSE)} is perhaps the most popular metric used in evaluating the accuracy of predicted ratings. The system generates predicted ratings $\hat r_{u,a}$ for a test set $T$ of user-advert pairs (u,a) for which the true ratings $r_{u,a}$ are known. The RMSE between the predicted and actual ratings is given by:
\begin{equation}\label{eq:RMSE}
	RMSE = \sqrt{\frac{1}{|T|}  \sum_{(u,a) \in T} {(\hat r_{u,a} - r_{u,a})}^2  }
\end{equation}
\textbf{Mean square error (MSE)} is an alternative version of RMSE, the main difference between these two estimators is simply that MSE has the same units of measurement as the square of the quantity being estimated, while RMSE has the same units as the quantity being estimated. \\
 
\textbf{Mean Absolute Error (MAE)} is a popular alternative, given by
\begin{equation}\label{eq:MAE}
	MAE = \sqrt{\frac{1}{|T|}  \sum_{(u,a) \in T} |{\hat r_{u,a} - r_{u,a}}|  }
\end{equation}

\subsection{Ad Click Prediction}\label{sec:}\vspace{-0.08cm}
In many applications the recommendation system tries to recommend adverts to users in which they may be interested. For example, when items are added to the queue, Amazon suggests a set of adverts on which the user would most probably click. In this case, we are not interested in whether the system properly predicts the ratings of these adverts but rather whether the system properly predicts that the user will click on them (e.g. they perform a conversion). Therefore, we then have four possible outcomes for a recommended advertisement, as shown in Table ~\ref{tab:table1}.

\begin{table}[!ht]
\small
\centering
\begin{tabular}{ l | c | c  }
&   Recommended & Not recommended  \\\hline
 Clicked & True-Positive (tp) & False-Negative (fn) \\
 Not clicked & False-Positive (fp) & True-Negative (tn)\\
\end{tabular} 
\caption{Classification of the possible result of a recommendation of an advert to a user}
\label{tab:table1}
\end{table} 
We can count the number of examples that fall into each cell in the table and compute the following quantities: \textbf{Precision} = $\frac{\#tp}{\#tp + \#fp}$, \textbf{Recall (True Positive Rate)} = $\frac{\#tp}{\#tp + \#fn}$, and \textbf{False Positive Rate (1 - Specificity)} = $\frac{\#fp}{\#fp + \#tn}$.

We can expect a trade-off between these quantities; while allowing longer recommendation lists typically improves recall, it is also likely to reduce the precision. We can compute curves comparing precision to recall, or true positive rate to false positive rate. Curves of the former type are known simply as precision-recall curves, while those of the latter type are known as a Receiver Operating Characteristic or ROC curves. A widely used measurement that summarizes the precision recall of ROC curve is the \textbf{Area Under the ROC Curve (AUC)}~\cite{BAMBER1975} which is useful for comparing algorithms independently of application.

\section{Prediction Experiments}\label{sec:results}\vspace{-0.08cm}

In this section we show results obtained for the two types of scenarios introduced in Sec.~\ref{sec:eval}. We conducted rigorous experiments to explore the strengths and weakness of the proposed algorithms when taking into account personality traits as features.

Let us say $X = \{ \vec x_1, ..., \vec x_N \}$ is the set of observations, where the vectors $\vec x_i$ correspond to features coming only from the group ``users' preferences'' as described in Table~\ref{tab:features}, and $N = 120$ stands for the number of users involved in the experiment. Regression is performed over the $20$ product categories. Given the user $u_i$, labels are assigned to each category by averaging the votes they gave to the category items such as $u_i =\{l_1,...,l_{20} \}, l \in [1-5]$. The prediction problem is solved using LR, L2-SVR, and SVR-rbf, while feeding them with and without features coming from ``personality traits''. All experiments were performed using a k-fold approach (k = $10$) and the folds are the same for every algorithm in comparison. In k-fold cross-validation, $X$ is randomly partitioned into k's equal sized subsamples. Of the k subsamples, a single subsample is retained as the validation data for testing the model, and the remaining k - 1 subsamples are used as training data. The cross-validation process is then repeated k times, with each of the k subsamples used only once as the validation data. The k results from the folds can then be averaged to produce a single estimation. 

The representation above serves as a basis for the feature ranking and selection strategy. Ranking features allow us to detect a subset of cues which is relevant and not redundant. Accordingly, we use the training data obtained after the split as input of the infinite feature selection (Inf-FS)~\cite{Roffo:InfFS:2015} algorithm. By construction, the Inf-FS is a graph-based method which exploits the convergence properties of the power series of matrices to evaluate the relevance of a feature with respect to all the other ones taken together. Indeed, in the Inf-FS formulation, each feature is mapped on an affinity graph, where nodes represent features, and weighted edges the relationships between them. In particular, the graph is weighted according to a function which takes into account both correlations and standard deviations between feature distributions. Each path of a certain length $l$ over the graph is seen as a possible selection of features. Therefore, varying these paths and letting them tend to an infinite number permits the investigation of the importance of each possible subset of features. The Inf-FS assigns a final score to each feature of the initial set; where the score is related to how much the given feature is a good candidate regarding the regression task. Therefore, ranking the outcome of the Inf-FS in descendant order allows us to perform the subset feature selection throughout a \textit{model selection stage}. In this way, we reduce the number of features, by selecting 75\% of the total. The selected features are: the number of favorite websites, T.V. programmes, sports, past times, the most watched movies and most visited websites, where we add the big-five personality traits. 
\begin{figure*}[!ht]\vspace{-0.18cm}
  \centering
    \includegraphics[width=1\textwidth]{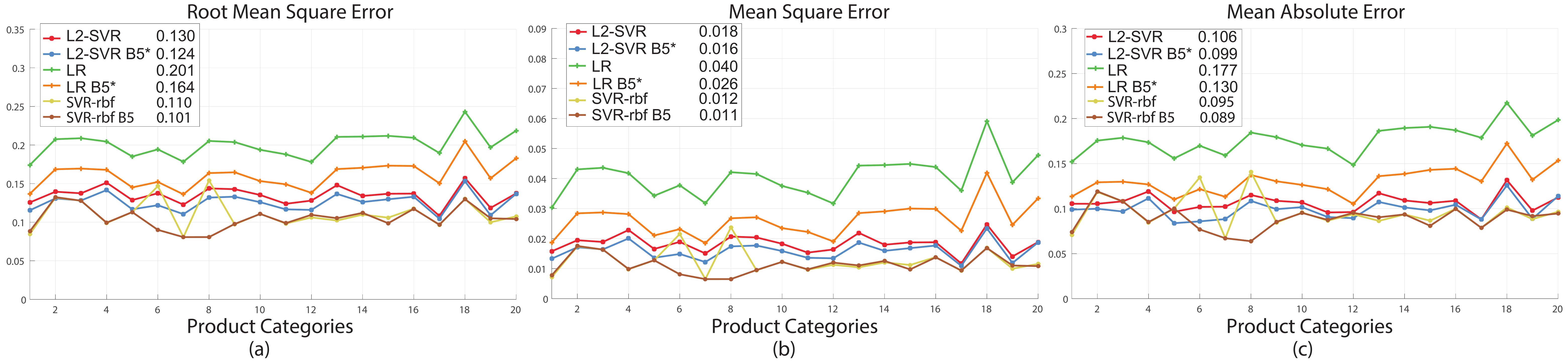}
 \caption{Measuring ratings prediction accuracy: B5 stands for Big-Five features. We indicate with an asterisk each method where B5 features, embedded into a baseline learner, shows a statistical significant effect over the baseline.}
\label{Figure:1}
\end{figure*}

\subsection{Exp.1 Ad Rating Prediction}\label{sec:}\vspace{-0.08cm}

In this section we report results for rating prediction showing that traces of user's personality can improve the prediction performance of the evaluated methods significantly. Statistical evaluation of experimental results has been considered an essential part of validation of machine learning methods. 

Figure~\ref{Figure:1} shows prediction results in term of RMSE, MSE and MAE. This first analysis shows how personality traits affect prediction performance. To this end, t-tests have been used for comparing prediction accuracies, showing a statistical significant effect of personality traits while using L2-SVR (p-value < 0.05) and LR (p-value < 0.01). 

\subsection{Exp.2 Ad Click Prediction}\label{sec:click}\vspace{-0.08cm}

This section shows an offline evaluation of click prediction. Along the lines of the previous experiment, a k-fold cross-validation is used. The prediction is performed by category, whenever a user showed their interest (clicks) on the majority of the items of a given category (50\% + 1) we labeled the category as ``clicked'', otherwise ``not clicked''. Therefore, given the methods in comparison, we compute precision-recall and ROC curves for each fold and for each category, and then we average the resulting curves over folds and categories. 
\begin{figure}[!ht]\vspace{-0.28cm}
  \centering
    \includegraphics[width=0.46\textwidth]{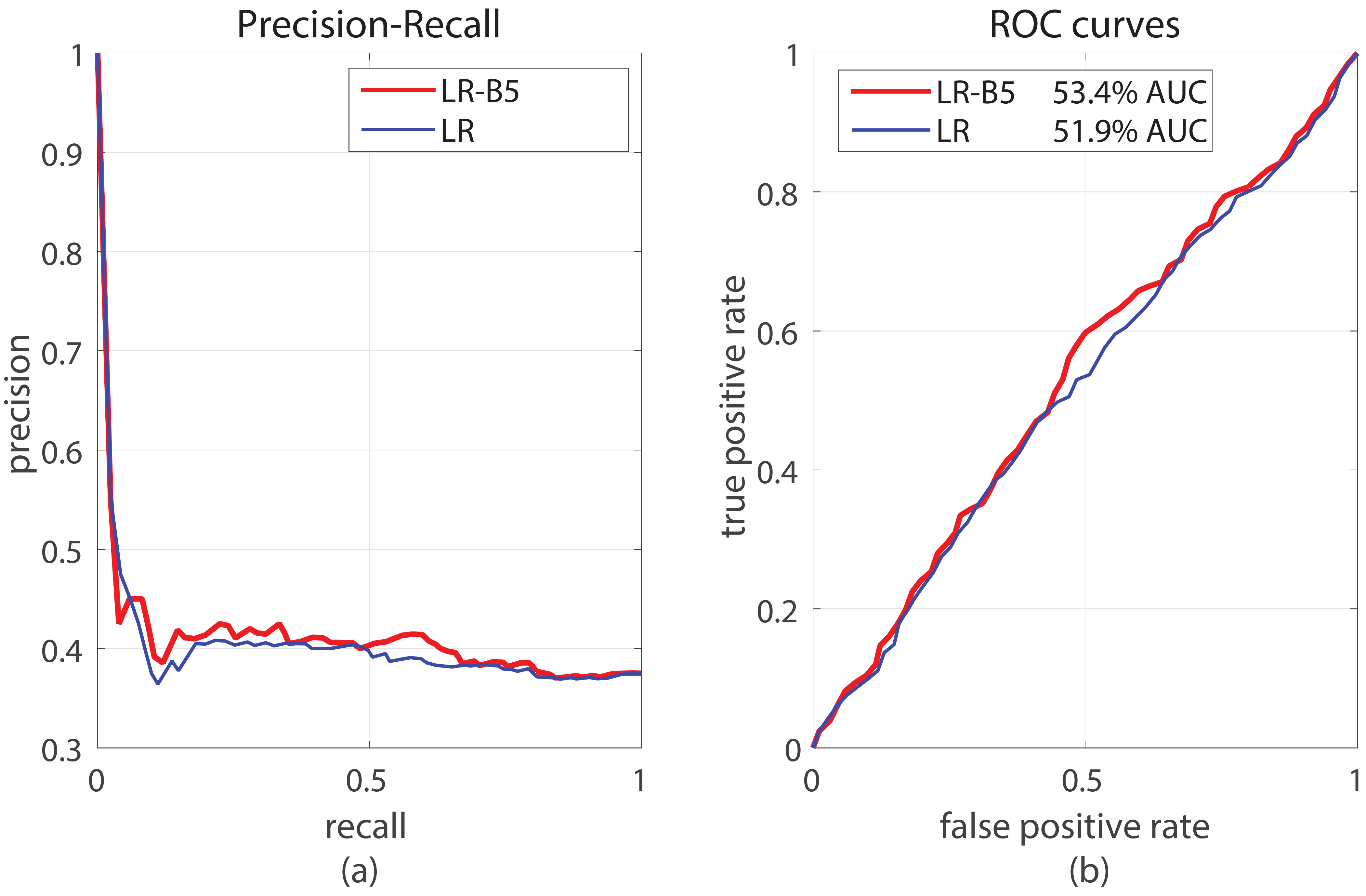}
 \caption{Comparison between LR and LR-B5: Curves show the proportion of preferred items that are actually recommended.}
\label{Figure:2}
\end{figure}
Figure~\ref{Figure:2}.(a) reports the precision-recall curves which emphasize the proportion of recommended items that are preferred and recommended. Figure~\ref{Figure:2}.(b) shows the global ROC curves for LR and LR-B5, which emphasize the proportion of adverts that are not clicked but end up being recommended. The LR-B5 curve completely dominates the other curve, the decision about the superior setting for LR is easy. 
\begin{table}[!ht]\vspace{-0.18cm}
\small
\centering
\begin{tabular}{ | l | l | l | l | }
\hline
\textbf{Method} & \textbf{ROC-AUC} & \textbf{Precision} & \textbf{Recall} \\\hline
L2-SVR        & 50.5\% & 39.2\% & 50.2\% \\\hline
\textbf{L2-SVR B5}   & 51.4\% & 39.9\% & 50.9\% \\\hline
LR                  & 51.9\% & 40.3\% & 51.3\% \\\hline
\textbf{LR-B5*}& \textbf{53.4\%} & \textbf{41.2\%} & \textbf{52.1\%}  \\\hline
SVR-rbf        & 48.3\% & 36.5\% & 48.8\%  \\\hline
\textbf{SVR-rb B5}    & 50.1\% & 38.2\% & 50.2\%\\\hline
\end{tabular} 
\caption{Performance for ad click prediction: The asterisk indicate that LR-B5 overcomes its baseline without Big-Five features. }
\label{tab:table2}
\end{table}
The Area Under the ROC Curve is calculated as a measure of accuracy, which summarizes the precision recall of ROC curves, we report AUC, precision and recall for all the methods in Table~\ref{tab:table2}. 

\section{Conclusions}\label{sec:conclusion}\vspace{-0.08cm}

In this paper, we presented the ADS Dataset, a collection of 300 real advertisements rated by 120 unacquainted individuals. The corpus has been collected with the main goal of studying the possible achievable benefits of employing personality traits in modern recommender systems. To obtain stronger and more relevant results for this community, appropriate and high-level features needed to be designed that carry important information for inference. In this paper, we only use raw data and mainly focus on the standard techniques used for recommendation of items in order to set a baseline for future work. We reviewed a large set of properties, and explain how to evaluate systems given relevant properties. We then discuss how to compare ARS based on a set of properties that are relevant for the application. Therefore, we review two main types of experiments in an off-line setting, where recommendation approaches are compared with different selections of features (i.e., with and without personality traits) accordingly with our goal. 
 
%
\bibliographystyle{abbrv}
\bibliography{sigproc}  

\begin{thebibliography}{10}

\bibitem{BAMBER1975}
D.~Bamber.
\newblock The area above the ordinal dominance graph and the area below the
  receiver operating characteristic graph.
\newblock {\em Journal of Mathematical Psychology}, 1975.

\bibitem{Bosnjak2007}
M.~Bosnjak, M.~Galesic, and T.~Tuten.
\newblock Personality determinants of online shopping: Explaining online
  purchase intentions using a hierarchical approach.
\newblock {\em Journal of Business Research}, 2007.

\bibitem{Chang:2011}
C.-C. Chang and C.-J. Lin.
\newblock Libsvm: A library for support vector machines.
\newblock {\em ACM Trans. Intell. Syst. Technol.}, 2011.

\bibitem{Chen201657}
J.~V. Chen, B.~chiuan Su, and A.~E. Widjaja.
\newblock Facebook c2c social commerce: A study of online impulse buying.
\newblock {\em Decision Support Systems}, 2016.

\bibitem{LIBLINEAR}
R.-E. Fan, K.-W. Chang, C.-J. Hsieh, X.-R. Wang, and C.-J. Lin.
\newblock {LIBLINEAR}: A library for large linear classification.
\newblock {\em Journal of Machine Learning Research}, 2008.

\bibitem{soBob}
B.~M. {Fennis} and A.~T. {Pruyn}.
\newblock You are what you wear: Brand personality influences on consumer
  impression formation.
\newblock {\em Journal of Business Research}, 2007.

\bibitem{Forrester}
Forrester.
\newblock Online retail industry in the us will be worth \$279 billion in 2015.
\newblock {\em TechCrunch}, February 28.

\bibitem{He:2014}
X.~He.
\newblock et al. practical lessons from predicting clicks on ads at facebook.
\newblock In {\em Data Mining for Online Advertising}, New York, NY, USA, 2014.
  ACM.

\bibitem{Kim:2011}
C.~Kim, K.~Kwon, and W.~Chang.
\newblock How to measure the effectiveness of online advertising in online
  marketplaces.
\newblock {\em Expert Syst. Appl.}, 2011.

\bibitem{Mowen2000}
J.~C. Mowen.
\newblock {\em The 3M Model of Motivation and Personality: Theory and Empirical
  Applications to Consumer Behavior}.
\newblock Springer US, Boston, MA, 2000.

\bibitem{NzVald20121186}
N.~{n}ez Vald{\'e}z.
\newblock et al. implicit feedback techniques on recommender systems applied to
  electronic books.
\newblock {\em Comput. Hum. Behav.}, 2012.

\bibitem{Avila16}
A.~C. P. and V.~M. S.
\newblock Click through rate prediction for display advertisement.
\newblock {\em International Journal of Computer Applications}, 2016.

\bibitem{rammstedt:2007}
B.~Rammstedt and O.~P. John.
\newblock {Measuring personality in one minute or less: A 10-item short version
  of the Big Five Inventory in English and German}.
\newblock {\em Journal of Research in Personality}, 2007.

\bibitem{Roffo_2013_ICCV_Workshops}
G.~Roffo, M.~Cristani, L.~Bazzani, H.~Quang~Minh, and V.~Murino.
\newblock Trusting skype: Learning the way people chat for fast user
  recognition and verification.
\newblock In {\em The IEEE International Conference on Computer Vision (ICCV)
  Workshops}, June 2013.

\bibitem{Roffo:HBU2014}
G.~Roffo, C.~Giorgetta, R.~Ferrario, and M.~Cristani.
\newblock {\em Just the Way You Chat: Linking Personality, Style and
  Recognizability in Chats}, pages 30--41.
\newblock Springer International Publishing, 2014.

\bibitem{Roffo:icmi2014}
G.~Roffo, C.~Giorgetta, R.~Ferrario, W.~Riviera, and M.~Cristani.
\newblock Statistical analysis of personality and identity in chats using a
  keylogging platform.
\newblock In {\em International Conference on Multimodal Interaction}. ACM,
  2014.

\bibitem{Roffo:InfFS:2015}
G.~Roffo, S.~Melzi, and M.~Cristani.
\newblock Infinite feature selection.
\newblock In {\em IEEE International Conference on Computer Vision (ICCV)},
  2015.

\bibitem{Turkyilmaz201598}
C.~A. Turkyilmaz, S.~Erdem, and A.~Uslu.
\newblock The effects of personality traits and website quality on online
  impulse buying.
\newblock 2015.
\newblock International Conference on Strategic Innovative Marketing.

\bibitem{wang2010click}
X.~Wang, W.~Li, Y.~Cui, R.~Zhang, and J.~Mao.
\newblock Click-through rate estimation for rare events in online advertising.
\newblock {\em Online Multimedia Advertising: Techniques and Technologies},
  2010.

\end{thebibliography}
%
%

\end{document}